\documentclass[letterpaper,10pt]{article}   
\usepackage{osajnl}   
 \usepackage{amssymb} 

\def\AOM {acousto-optic modulator}

\def\ECDL {extended cavity diode laser}
\def\EDFA {Er-doped fibre amplifier}

\def\HNLF {highly nonlinear fibre}
\def\MOT {magneto-optical trap}

\def\ppKTP {periodically poled potassium titanyl phosphate}  

\def\PMC {photo-multiplier cell}

\def\SAS{saturated absorption spectroscopy}

\def\snr {signal-to-noise ratio}

\def\wrt {with respect to}

\def\ARC {Australian Research Council}

\newcommand{\degC}{$^{\circ}$C}

\newcommand{\si}{$\sim$}
\newcommand{\um}{$\mu$m}
\newcommand{\uK}{$\mu$K}  
\newcommand{\uW}{$\mu$W}

\newcommand{\fastT}{$^{1}S_{0}-\,^{1}P_{1}$}    
\newcommand{\coolingT}{$^{1}S_{0}-\,^{3}P_{1}$}  

\newcommand{\clockT}{$^{1}S_{0}-\,^{3}P_{0}$}

\newcommand{\clockTdash}{$^{1}S_{0} -\,^{3}P_{0}$} 
\newcommand{\diffP}{$^{3}P_{0} -\,^{3}P_{1}$}

\newcommand{\Yb}{$^{171}$Yb}	

\newcommand{\Ybthree}{$^{173}$Yb}

\begin{document}

\title{Clock and inter-combination line frequency separation   in $^{171}$Yb } 


\author{L. Nenadovi\'c  and J.J. McFerran$^{*}$}  
\address{School of Physics, University of Western Australia, 6009 Crawley, Australia}
\address{$^*$Corresponding author: john.mcferran@uwa.edu.au}

\begin{abstract}

We have carried out concurrent optical frequency measurements of the $(6s^{2})\,^{1}S_{0} -(6s6p)\,^{3}P_{1}$ ($F=\frac{3}{2}$) and $(6s^{2})\,^{1}S_{0} -(6s6p)\,^{3}P_{0}$ transitions in $^{171}$Yb, and in so doing have determined the frequency separation between the $^{3}P_{0}$ ($F=\frac{1}{2}$) and  $^{3}P_{1}$ ($F=\frac{3}{2}$) levels with an uncertainty in the ppb range.   The knowledge of this frequency interval will aid experiments relying on the $^{1}S_{0}-\,^{3}P_{0}$ line in $^{171}$Yb and $^{173}$Yb  without access to highly accurate frequency standards. 

\end{abstract}

                           
                            



\maketitle 

\section{Introduction}

Neutral ytterbium is used heavily in cold and ultra-cold atomic and molecular investigations~\cite{Tak2003b,Fuk2007,Fuk2007a}.  
  Increasingly the spin-and-dipole forbidden transition [$(4f^{14}6s^{2})\,^{1}S_{0} -(4f^{14}6s6p)\,^{3}P_{0}$]  in the composite fermionic  Yb isotopes is being exploited; from applications in atomic clocks~\cite{Hin2013,Lem2009a,Yas2012,Par2013,Tak2015a,Piz2012,Nin2013},  to explorations in quantum many body interactions~\cite{Ger2010a,Dal2011,Pag2014,Sca2014,Non2013}.  
  Optical lattice clocks based on ytterbium are becoming more common as their performance continues to improve, whether it be accuracy~\cite{Lem2009a,Yas2012,Par2013, Tak2015a}, stability~\cite{Hin2013, Tak2015a}, or reproducibility~\cite{Lem2009a,Yas2012,Par2013, Tak2015a}.   However, before an atomic clock can become a versatile frequency reference, it is helpful if the transition can be easily instituted 
  without relying upon substantial infrastructure, such as Cs primary  frequency standards or hydrogen masers.  Since the ensemble of Yb \coolingT\ isotopic lines provides a relatively easy means of access,  a comparison between the  \coolingT\ and the \clockT\  lines of \Yb\ affords a convenient means of compressing the search space when exploring for the clock transition. 
  Here we report on the frequency  difference  between the $^{1}S_{0}\, (F=\frac{1}{2})-\,^{3}P_{1}\, (F=\frac{3}{2})$ and $^{1}S_{0}\, (F=\frac{1}{2})-\,^{3}P_{0}\, (F=\frac{1}{2})$  transitions in \Yb\ with a $1-\sigma$ uncertainty below a part in $10^{8}$. 
  Our measurements will also assist those searching for the \clockT\ line(s) in \Ybthree, given the previously reported value for its transition frequency~\cite{Hoy2005}.      Furthermore, the absolute optical frequencies of the other \coolingT\  isotopic lines can be inferred from previously reported measurements~\cite{Cla1979,Wij1994,Pan2009}. 

The \clockT\ line in fermonic Yb is finding a purpose in many applications. Apart from clocks, it is being used to explore quantum many body interactions~\cite{Ger2010a,Dal2011,Pag2014,Sca2014,Non2013}, and has been proposed for use in quantum simulation studies~\cite{Dal2008, Shi2009,Dal2011a} and gravitational inverse-square-law tests at micron scales~\cite{Wol2007}. 
The natural line-width of the  clock (\clockT) transition is extremely narrow, approximately 8\,mHz.   The line-width can of course be intensity broadened by the probe light, though, this is more easily carried out through the use of ground state broadening; for example, by coupling light to the \fastT\ or \coolingT\ transitions.  This 
method can produce line-widths of a few MHz, but  fractionally this is still only a few ppb.   Hence, finding the spectroscopic line without access to an accurate frequency reference can pose a challenge. 
The above experiments using Yb can be greatly simplified if the expensive and/or intricate frequency standard becomes inessential.  
With the \diffP\ or \coolingT\ frequencies reported here, a wave-meter with sufficiently high resolution (and $\leq10^{-6}$ level accuracy) should be the only frequency measuring device 
needed to find  the \clockT\ lines in \Yb\ or \Ybthree. 
There are two hyperfine levels in the $^{3}P_{1}$ state of \Yb\ with $F=\frac{1}{2}$ and $F=\frac{3}{2}$.  Our measurements focus on the $^{1}S_{0}\, (F=\frac{1}{2})-\,^{3}P_{1}\, (F=\frac{3}{2})$ transition, which is commonly used for laser cooling. Henceforth, denoted with \coolingT.  Based on an absolute frequency measurement of the \coolingT\ transition in $^{176}$Yb and the relevant isotopic shift~\cite{Pan2009}, a previous estimate for the $^{1}S_{0}\,-\,^{3}P_{1}\, (F=\frac{3}{2})$ transition frequency is $539\,390\,366\,\pm10$\,MHz.  We find the frequency to be offset from this by  nearly four standard deviations. 

In our experiment, the clock transition interaction is carried out with \Yb\ atoms held in a magneto-optical trap, while the \coolingT\ transition is addressed via  saturation absorption spectroscopy on a thermal beam.   
In the following we present details about  (i) the optical cavity used to stabilise the laser light that is needed to probe the clock transition, (ii) the spectroscopic methods for producing the \coolingT\ and \clockT\  lines, (iii) the frequency measurement scheme, (iv) and a discussion of results including systematic frequency shifts. 

\section{Experimental setup}

\begin{figure}[h]
 \begin{center}
{		
  \includegraphics[width=9cm,keepaspectratio=true]{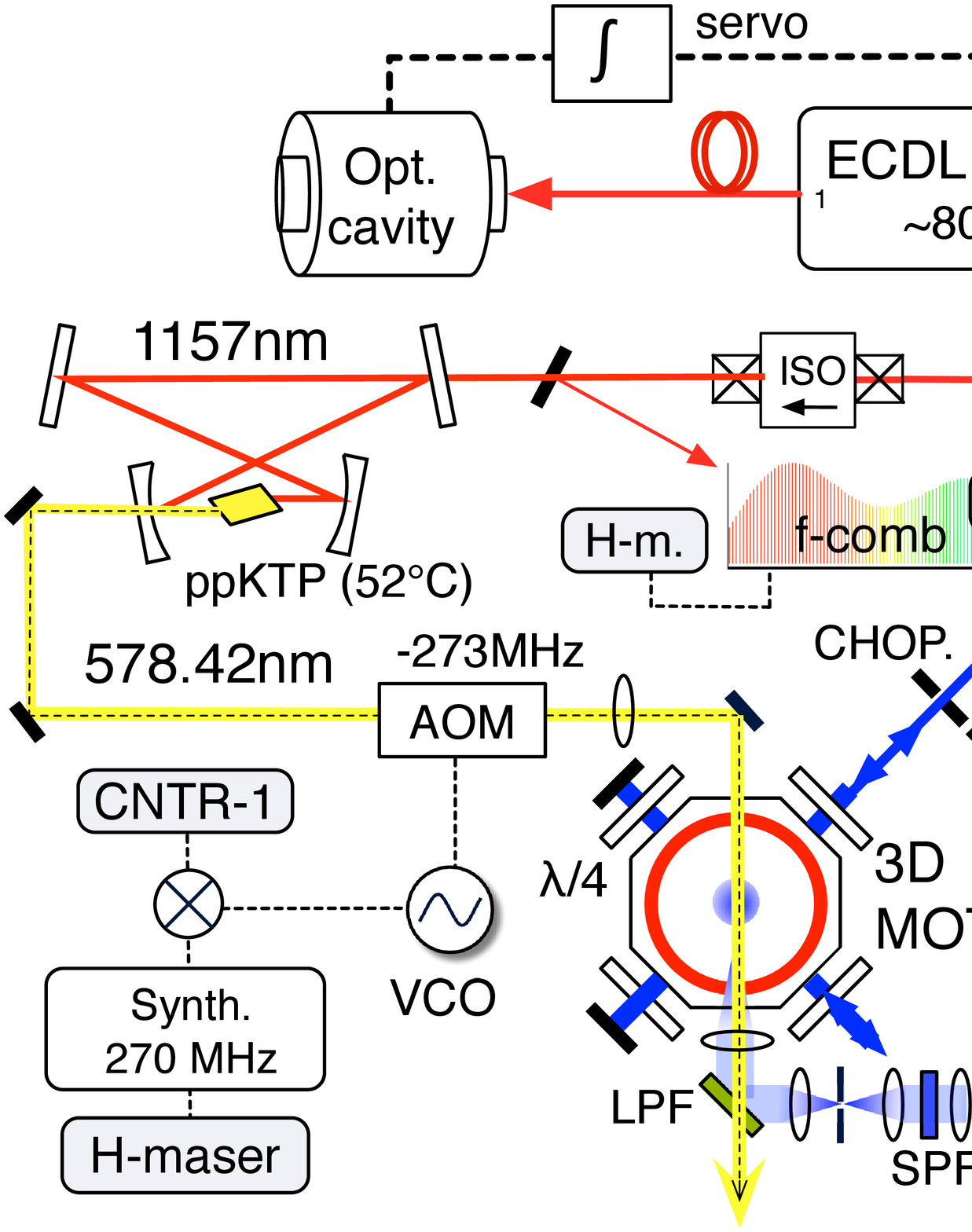}}   
\caption[]{\footnotesize     
Experimental arrangement for the \clockT\ line spectroscopy in \Yb. In some measurements the chopper wheel is used to shutter the 398.9\,nm MOT light, in others, the two AOMs are used in combination with each other.  AOM, \AOM; CNTR, frequency counter; ECDL, \ECDL; f-comb, frequency comb (in the near-IR); H-m, hydrogen maser; I, current modulation port; ISO, optical isolator; LPF, long-wavelength pass filter; MOT, \MOT; PMC, \PMC; PPKTP, \ppKTP; PZT, piezo transducer; SPF, short-wavelength pass filter; VCO, voltage controlled oscillator.
  }
   \label{ClockSpectroscopyWithAOM}  
\end{center}
\end{figure}

Ytterbium atoms are maintained in a 3D \MOT\ fed with a zero-$B$-field crossing  Zeeman slower~\cite{Kos2014}. 
  Laser light at 398.9\,nm, for the MOT and Zeeman slower, is generated via resonant frequency doubling of 797.8\,nm light sourced from a Ti:sapphire laser.   The approximate wavelengths of the  \coolingT\ and \clockT\ transitions are 555.8\,nm and 578.4\,nm, respectively.  Separate laser schemes have been put in place  to address the two transitions; both make use of near-infrared lasers that are resonantly frequency doubled.  
The principal components and layout for the clock transition interaction and measurement are shown in Fig.~\ref{ClockSpectroscopyWithAOM}. 
To probe the \clockT\ we commence with a semiconductor laser in an extended cavity at 1157\,nm (Time-Base GmbH) and resonantly frequency double the light with 
 a \ppKTP\ (PPKTP) crystal as the nonlinear element. 
 The poling period of the crystal is 11.6\,\um\ and its temperature is maintained at 51.8\,\degC. 
 We use the same  set of optics for the resonant cavity as those used for the 1112\,nm frequency doubling~\cite{Kos2015}.  The mode-matching of the 1157\,nm light into the cavity has been partly optimised;  sufficient to generate 3\,mW of 578.4\,nm light with 17\,mW of IR light incident. 
 To avoid modulation of the 578.4\,nm light, but still maintain laser  locking to the centre of a frequency doubling cavity resonance, we employ the H\"{a}nsch-Couillaud frequency stabilisation scheme~\cite{Han1980}. 
 
\subsection{Optical cavity and frequency comb}

To stabilise the frequency of the 1157\,nm light we use an optical cavity  with a spacer composed of ultra-low expansion (ULE) glass and  fused silica mirror substrates (Stable Lasers Systems and Advanced Thin Films).  The finesse at 1157\,nm is  $400(20)\times10^{3}$, based upon an average of cavity ring-down measurements.  One mirror is planar and the second has a radius of curvature of 500\,mm, the former we use  as the input coupler.  The spacer is cylindrical, where both the length and diameter of the spacer are 100\,mm in dimension.    Two notches, cut either side,  run the length of the cylinder,  providing  a support plane 3.6\,mm below the mid plane. 
Four symmetrically arranged Viton pads separate the ULE spacer from a Zerodur base.
The Zerodur mounted optical cavity is enclosed by a gold-coated aluminium shield and  placed inside a vacuum enclosure.  
  Surrounding the vacuum housing are two nickel coated aluminium shells, 9\,mm thick, with space for thermal insulation between each. 
  The vacuum chamber is actively temperature stabilised, while the outer shells provide passive stabilisation (there is also the option to stabilise the  inner aluminium shell, but this is not used at present).  The complete ensemble is placed on an active vibration platform and inside an acoustic-damping enclosure.   The thermal time constant between the  temperature  control stage and the optical cavity is \si20 hours, and that between the outer shell and thermal control stage is 16\,hours.   The pressure in the vacuum chamber is maintained at \si$1.0\times10^{-7}$\,mbar. 
  
  We employ the Pound-Drever-Hall scheme  to stabilise the 1157\,nm laser to the optical cavity.  For phase modulation we use a resonant electro-optic modulator driven at 20\,MHz with a modulation depth of 5-10\,\%.  Approximately  110\,\uW\ of light is sent to the cavity with 75-80\,\% coupled into the lowest order Gaussian-spherical mode. The correction signal is fed back to current modulation on the 1157\,nm ECDL, and is integrated, producing slower feed-back to a piezo transducer controlling the length of the laser cavity.  The servo bandwidth on the current modulation is \si320\,kHz~\cite{note1}. 
  

  The optical cavity's frequency dependence on temperature has been examined by comparing the frequency of the cavity-stabilised 1157\,nm laser against a hydrogen-maser, via a frequency comb. To generate the frequency comb light at 1157\,nm (and 1112\,nm), we amplified the output of a MenloSystems 1.55\,\um\ fibre oscillator with an Er-doped fibre amplifier, before coupling into 90\,cm of highly non-linear fibre (OFS Fitel). 
 One metre of   Er single mode fibre is used in the amplifier, which has   30\,dB\,m$^{-1}$ absorption at 1530\,nm, and the nonlinear coefficient of the \HNLF\ (HNLF) is 11.5\,W$^{-1}$\,km$^{-1}$ with a dispersion parameter, $D$, of +3\,ps\,nm$^{-1}$\,km$^{-1}$~\cite{Nic2003}. %
Dispersion management  with dispersion compensated fibre and single mode fibre (SMF28) is used between the \EDFA\ and the HNLF.    Direct splicing between the SMF28 and the HNLF produced a  loss of \si0.6\,dB.  

The mode-spacing of the frequency comb is stabilised with a dielectric resonant oscillator, which  is phase referenced to a hydrogen-maser (Kvarz CH1-75A).   Heterodyning the 1157\,nm light and an adjacent comb element on a photodiode provides an RF beat signal for frequency counting.  
This  beat frequency is measured for different cavity temperatures  once equilibrium is reached (i.e. after at least 3 days), and to assess the temperature, 
  recordings are taken from  points on the
     vacuum enclosure.  
   We have determined  the temperature at which the coefficient of thermal expansion is zero ($T_{CTE=0}$) to be $29.4\pm 0.7$\,\degC, and the second order thermal expansion coefficient is \si $1.2\times10^{-9}$\,K$^{-2}$.   Having the cavity temperature set to within 0.9\,\degC\ of  $T_{CTE}=0$  should be sufficient to have the frequency instability of the locked laser limited by the thermo-elastic noise of the fused silica mirror substrates~\cite{Num2004, Not2006}. We estimate the temperature variations to be $\lesssim$100\,nK at 1\,s at the cavity based on temperature measurements of the vacuum can and passive low pass filtering between the shield and the cavity.   

\begin{figure}[h]
 \begin{center}
{		
  \includegraphics[width=12cm,keepaspectratio=true]{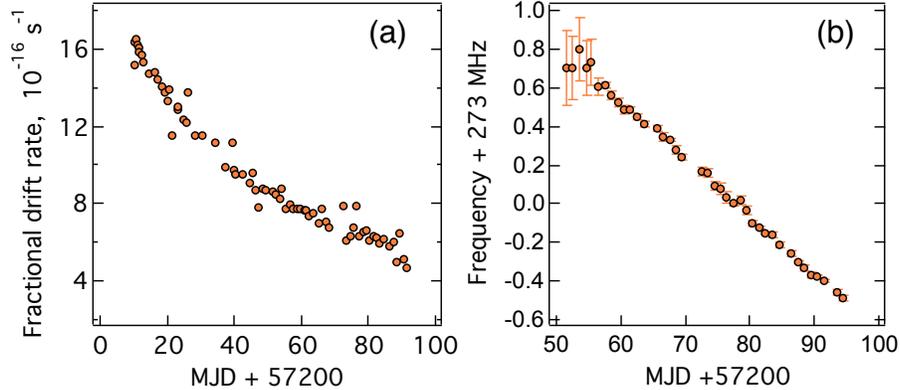}}   
\caption[]{\footnotesize     
(a) The fractional frequency drift rate between the optical-cavity-stabilised 1157\,nm laser and a hydrogen maser.     (b) The frequency difference between the cavity stabilised 578.4\,nm laser (stabilised at 1157\,nm) and the \clockT\ transition. The frequency shown is that driving an \AOM\ used to bridge the frequency difference (offset by 273.0\,MHz).  
 }
   \label{Drift}  %
\end{center}
\end{figure}

We measure the rate of frequency drift between the H-maser and the optical cavity via the frequency comb by tracking the measured beat frequency from day to day.  These measurements are shown in Fig.~\ref{Drift}(a).  We observe a very slow decline in the drift rate over a monthly time scale, which has now reached  \si150\,mHz\,s$^{-1}$  ($6\times10^{-16}$\,s$^{-1}$).    
 The reason for the monotonic drift is not completely certain. One may posit that it is due to creep in the ULE spacer; however, other optical cavities have shown that this effect is nearly an order of magnitude smaller~\cite{McF2012b}.  The frequency drift of the cavity-stabilised-laser \wrt\ the \clockT\ line can also be monitored and is discussed  below. 

 \subsection{Spectroscopy methods}

We perform spectroscopy on the \clockT\ transition with the Yb atoms maintained in a \MOT\ that uses the \fastT\ line at 398.9\,nm.   A description of the MOT is found in Refs. \cite{Kos2014,Kos2015}.
Initially, the search for the clock line was carried out by locking the 1157\,nm laser  to an adjacent mode of the frequency comb and scanning the mode spacing  (i.e. repetition rate of the mode-locked laser that seeds the comb generation).   To determine the approximate frequency of the 1157\, nm light prior up to the detection of the clock transition we  used the H-maser-steered frequency comb.  An estimation of the comb element number, $N$ (in the IR), was made by stepping the mode spacing by 950\,Hz and observing the frequency step in the optical domain.   A measurement of the stepped beat note frequency over \si30 minutes led to an $N$ value of $1036592\pm1$.  Repeated measurements over several days reduced the uncertainty  to $0.8$. 
  This approach was necessary, since we were unaware of any 8-digit precision measurements of any of the \coolingT\ lines in the Yb isotopes.
  Alignment between the 578.4\,nm probe beam and the atoms was afforded through the use of a tracer beam tuned to the \coolingT\ transition (the tracer beam produced depletion of atoms from the MOT, signalling overlap).    
  Detection of the ground-state atoms is made by collecting 398.9\,nm fluorescence that is spatially filtered to minimise scattered light from the MOT beams.   Within the imaging scheme there is one long-wavelength pass filter below the chamber and two short-wavelength pass filters before a \PMC\ (PMC).  The cut-off wavelengths in each case are \si505\,nm  (this gives access to the 555.8\,nm and 578.4\,nm beams after they pass through the chamber).    
  
After the first detection of the \clockT\ line a change in the method for tuning the 578.4\,nm light frequency was made, where 
instead of locking  the 1157\,nm laser to the comb, 
it was locked directly to the optical cavity.   The comb enabled a broad frequency tuning range, but poses a limit on spectral resolution set by the multiplied noise of the dielectric resonant oscillator.
For frequency tuning,  the 578.4\,nm light is passed through an \AOM\ (AOM) with an RF drive frequency at \si273\,MHz, negatively shifted \wrt\ the optical cavity mode.  The drive frequency  is heterodyned with a 270\,MHz signal, referenced to the H-maser, and the frequency difference counted.  

 The \clockT\ transition is detected by three different means. The simplest approach uses rapid switching of the 398.9\,nm MOT beams by use of an optical chopper wheel, while the 578\,nm light remains on continuously.   The photomultiplier signal  is sent to a locking amplifier (LIA) that uses the chopper frequency as its reference signal, and  the demodulated output of the LIA provides the spectroscopic signal.   This method  leaves the atoms prone to a light shift since the trapping  and probe light are on simultaneously for half the time. 
  A second approach uses interleaving of the 398.9\,nm and 578.4\,nm light using respective AOMs, where the on and off periods are equal and 400\,$\mu$s in duration.   By comparing the two methods, the line-centre frequency shifts upward by \si$3\times10^{-10}$  in the case of modulating the 398.9\,nm light only. 
 A third approach has also been used that further  quantifies the light shift (described below).

 \section{Results}
 
\begin{figure}[h]
 \begin{center}
{		
  \includegraphics[width=12cm,keepaspectratio=true]{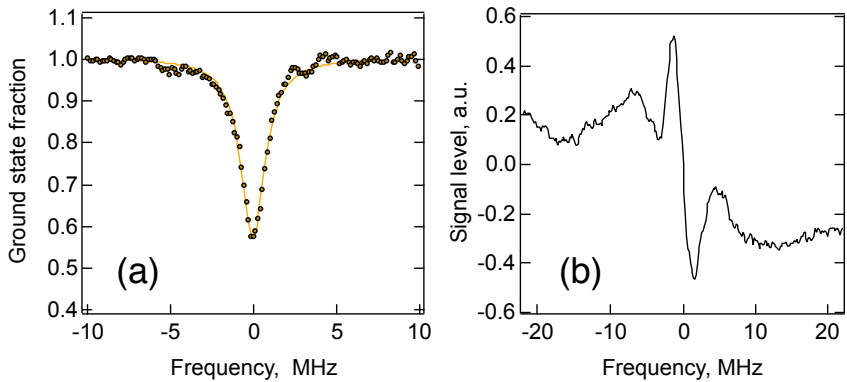}}   
\caption[]{\footnotesize     
(a) A spectrum of the \clockT\ transition through detection of 398.9\,nm fluorescence. Here the 398.9\,nm MOT beams were switched on and off at 2.2\,kHz with a rotating chopper wheel, while the 578.4\,nm probe beam was maintained continuously.    (b) The derivative spectrum produced from saturated absorption  of the $^{1}S_{0}\, (F=\frac{1}{2})-\,^{3}P_{1}\, (F=\frac{3}{2})$ transition in \Yb\ from a thermal beam where the magnetic substates remain degenerate.  
 }
   \label{Spectra}  %
\end{center}
\end{figure}

The majority of the clock-line spectroscopy was performed by  chopping the  398.9\,nm MOT beams at 2.2\,kHz, while leaving the 578.4\,nm probing acting continuously~\cite{Hon2005a}.  
The probe excites the atoms to the metastable $^{3}P_{0}$ level, causing a depletion from the ground-state, and hence a reduction in fluorescence reaching the photomultiplier cell. 
 With the modulated MOT beams, the temperature is not as cold as can be achieved with an optimised cyclic sequence.    However, for the purposes of measuring the \ \diffP\ frequency separation, the approach is adequate, since the accuracy is limited by the spectroscopy on the \coolingT\ line.  The LIA output signal and the AOM frequency are logged simultaneously. 
An example of the clock-line spectrum is shown in Fig.~\ref{Spectra}(a), where the level of atom number depletion reaches \si40\,\%. 
 A discussion relating to the line-width appears below. 

The drift of the optical cavity frequency can also be determined  by tracking the change in the AOM frequency that is used to tune the 578.4\,nm light to the \clockT\ transition (the AOM bridges the frequency interval between 2$f_{1157\mathrm{nm}}$ and the clock transition, where $f_{1157\mathrm{nm}}$ is the frequency of the 1157\,nm laser locked to the optical cavity).   In one measurement run the 578.4\,nm light is swept across the clock transition in both directions multiple times with a total duration of \si4\,min.  
 A Lorenztian line-shape is fit to  the signal versus frequency data, from which the centre frequency is determined. 
  This scheme is repeated at least five times and an ensemble average of the centre frequencies is made for each day's measurement.  The centre frequencies (that of the AOM) are shown in Fig.~\ref{Drift}(b).  The mean drift rate over the last 20 days is $6.4\times10^{-16}$\,s$^{-1}$, in close correspondence with the drift rate determined via the comparison with the H-maser.  In principle, a difference between these two drift rates should show the residual drift of the H-maser frequency (but we do not yet claim to be sensitive to this).

The \Yb\ \coolingT\ line was generated through saturation absorption spectroscopy (SAS)~\cite{Dem2008} on a thermal beam with \si50\,mrad angular divergence.  Approximately 280\,\uW\  of 555.8\,nm light passes through the atomic beam (the same thermal beam that leads to a Zeeman slower and the \MOT) at right angles, and is retro-reflected with a cat's eye that uses a lens-mirror combination (focal length =75\,mm).  The 555.8\,nm beam shape is elliptical with the major axis aligned with the flow of the atoms. The corresponding single-beam saturation parameter is $s_{0}\approx14$ (intensity broadening is evident, but the added intensity assists with the SNR).   
We apply a vertical bias magnetic field  to compensate for the Earth's $B$-field so that the magnetic sub-states   of the $^{1}S_{0}\, (F=\frac{1}{2})$  and $^{3}P_{1}\, (F=\frac{3}{2})$ levels remain degenerate.   
The polarisation of the light is  linear (and vertical), so that $\Delta m_{\mathrm{F}}=0$.  

 Prior to reaching the atoms the light is  positively frequency shifted by \si110\,MHz with an \AOM.  Frequency modulation spectroscopy is carried out by modulating the AOM-shifted light at 33\,kHz, sourced from another lock-in amplifier.  A second photomultiplier cell collects the 555.8\,nm fluorescence, producing an input signal for the LIA.  
  An example of the  discriminator signal produced by the  LIA demodulation (first harmonic) is shown in Fig.~\ref{Spectra}(b).   Most of the width (\si 2.9\,MHz) is caused by intensity broadening, with a small contribution from transit-time broadening (\si40\,kHz) and the natural line-width (184\,kHz~\cite{Bel2012a}).
The 555.8\,nm laser light is frequency stabilised   
by using the discriminator as a correction signal and feeding back to the 1112\,nm master laser via a piezo transducer with a \si1\,Hz bandwidth.   For line-centre measurements, the third harmonic from the lock-in amplifier is used to minimise the influence of the Doppler background. 

To perform the frequency evaluation for the \Yb\, \coolingT\ line we again create a heterodyne-beat in the near IR, this time mixing  1112\,nm single frequency light with a portion of the frequency comb.  Approximately 2\,mW of cw light is split from the  master laser and sent for comb mixing, leaving more than 2\,mW available  
 for the slave injection locking (the injection locking remains robust for injected power levels above 100\,\uW).     A beat \snr\ (SNR) of 20\,dB in a resolution bandwidth of 500\,kHz is sufficient for frequency counting.  A bandpass filter with 10\,MHz FWHM and centred at 30\,MHz is applied to the beat signal. 
 The comb element number is evaluated by use of a wave-meter with an accuracy of \si$4\times10^{-6}$.   
 Frequency counting of the 1112\,nm beat signal (with the comb) is carried out while the master laser is locked to the centre of the saturation dip.
Measurements have also been carried out by creating an error signal involving the 555.8\,nm light and the cold ytterbium atoms.  Similar frequency values are found; however, the additional laser cooling of the 555.8\,nm light  distorts the spectral line creating sizeable frequency shifts~\cite{Kos2015}.  

  \begin{figure}[h]
 \begin{center}
{		
  \includegraphics[width=12cm,keepaspectratio=true]{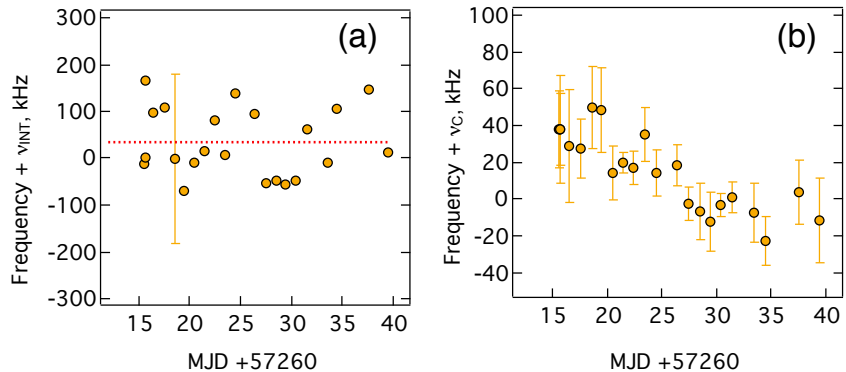}}   
\caption[]{\footnotesize     
Frequency measurements versus modified Julian day for the (a) \coolingT\ and (b) \clockT\ transitions in \Yb.  The frequencies are offset by $\nu_{\mathrm{INT}}=539\,390\,405\,400$\,kHz and $\nu_{\mathrm{C}}=518\,295\,836\,590.865$\,kHz~\cite{Lem2009a}, respectively.   The dotted line in (a) represents the statistical mean equal to $\nu_{\mathrm{INT}}+33$\,kHz and the error bars  represent the estimated systematic uncertainty (for all the data points). 
 }
   \label{LineMeasurements}  %
\end{center}
\end{figure}


To measure the frequency difference between the $^{3}P_{1}$ and $^{3}P_{0}$ levels we carry out separate measurements of the \clockT\ and \coolingT\ transition frequencies and take the difference.   The frequency determination is a two part process.  (i)   The near-IR frequencies of the lasers are determined using the frequency comb equation $f_{\mathrm{NIR}}=nf_{\mathrm{r}}\pm f_{\mathrm{o}} \pm f_{\mathrm{b}}$,  where $n$ is the number of the comb element accounting for the optical beat signal,  $f_{\mathrm{r}}$ is the mode spacing, $f_{\mathrm{o}}$ is the offset frequency and $f_{\mathrm{b}}$ is the beat frequency.  The signs of  $f_{\mathrm{b}}$ and  $f_{\mathrm{o}}$ are easily determined by the usual means.  For the measurements  of both transitions $f_{\mathrm{o}}=-20.0$\,MHz.  (ii)  The second part involves the spectroscopy, where the line-centre frequency, $ f_{\mathrm{AOM\_C}}$, is determined with the use of an AOM  (either by locking to the line and frequency counting, or by sweeping across the transition, then averaging and lineshape fitting).  Thus the resultant optical frequency  is  $f_{\mathrm{Opt}}=2f_{\mathrm{NIR}}\pm f_{\mathrm{AOM\_C}}$.   For the \coolingT\ and \clockT\ lines the AOM frequency shift is positive and negative, respectively.  The measurements of the \coolingT\ and \clockT\ transition frequencies are shown in Fig.~\ref{LineMeasurements}(a) and Fig.~\ref{LineMeasurements}(b), respectively.  For clarity we have offset the frequencies by  $\nu_{\mathrm{INT}}=539\,390\,405\,400$\,kHz in Fig.~\ref{LineMeasurements}(a) and $\nu_{\mathrm{C}}=518\,295\,836\,590.865$\,kHz~\cite{Lem2009a} in Fig.~\ref{LineMeasurements}(b).  Measurements for the clock transition agree with previously reported values to within $8\times10^{-11}$, and the remaining scatter  is mostly attributed to the uncertainties related to the 398.9\,nm light shift~\cite{note2}. 
These variations are small compared to the uncertainties of the \coolingT\ line-centre measurements, so do not impact the   \diffP\ frequency measurements greatly.
The error bars on the \clockT\ line frequencies are 1-$\sigma$ statistical uncertainties arising from the line-fitting procedure. The systematic shifts will be discussed shortly. 
  For the \coolingT\ transition the statistical uncertainty of any particular day's measurement is below 10\,kHz, but  over the duration of several weeks we see a greater variation.  
 Because it is  \SAS\ and we use third harmonic detection with the lock-in amplifier, the first order Doppler shifts should be  negligible. There remains a second order Doppler shift, but this  is below the $10^{-14}$ level (with a 50\,mrad atom beam divergence and $v_{\mathrm{mp}}\sim320$\,m\,s$^{-1}$).   
The main reason for the variations is that saturated absorption spectroscopy is wavefront curvature sensitive~\cite{Bor1976,Hal1976a}. 
To estimate a related systematic uncertainty we measured the line-centre frequency versus the cat's eye mirror position. Over a 300\,\um\ displacement in the beam propagation direction,  the shift is found to be less that 20\,kHz. 
 A more significant shift is observed for lateral displacements of the cat's eye lens.  Here the evaluation of  the optimum alignment leaves a $1-\sigma$ uncertainty of about 170\,kHz, where we compare beam size and overlap of the incident and return beams at 1\,m from the atoms. 
  We note that the symmetry of the derivative spectrum in Fig.~\ref{Spectra}(b) is high, implying that a systematic frequency shift due to wavefront curvature should be small in comparison to the line-width~\cite{Hal1976a}.
 A shift could also occur if there is a difference in intensity for the two beams (together with curved wavefronts)~\cite{Hal1976a}, but  the windows and lenses are antireflection coated and the mirror has a reflectivity of 99.5\,\% to minimise related shifts. 
 
 \begin{figure}[h]
 \begin{center}
{		
  \includegraphics[width=12cm,keepaspectratio=true]{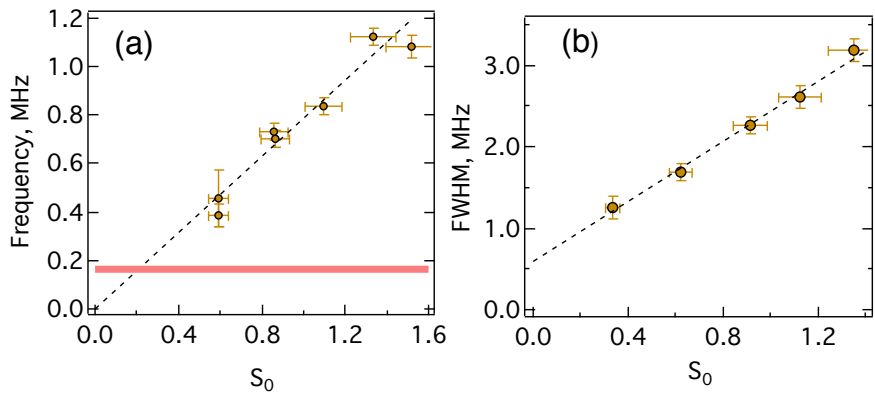}}   
\caption[]{\footnotesize     
(a) \clockT\  line-centre frequency versus the 398.9\,nm MOT light intensity during a probe time of 75\,ms. The horizontal line is the frequency measured by use of chopping the 398.9\,nm MOT light at 2.2\,kHz and having the 578.4\,nm light on continuously.   (b)   \clockT\ line-width (full-width half-maximum) versus 398.9\,nm intensity.  The saturation parameter $s_{0}=I/I_{\mathrm{sat}}$, where $I_{\mathrm{sat}}=595$\,W\,m$^{-2}$.
 }
   \label{LightShift}  %
\end{center}
\end{figure}

 We note that the $^{1}S_{0}\, (F=\frac{1}{2})-\,^{3}P_{1}\, (F=\frac{3}{2})$ transition in \Yb\ should be 2.7\,MHz away from the $^{1}S_{0}\, (F=\frac{5}{2})-\,^{3}P_{1}\, (F=\frac{3}{2})$ transition in \Ybthree\ at zero $B$-field, where the relative strength of the latter is about 30\,\% of the former~\cite{Pan2009}.  
Applying a dc magnetic bias field in our case  creates two dominant derivative lines in the \coolingT\ spectrum, consistent with the $m_{F}=-\frac{1}{2}\rightarrow m_{F'}=-\frac{1}{2}$ and  $m_{F}=\frac{1}{2}\rightarrow m_{F'}=\frac{1}{2}$ transitions in \Yb; hence, the influence of the \Ybthree\ line is only small and not easily detectable (for reasons not immediately clear).  However, it may still  cause a perturbation to the  \coolingT\ (\Yb) line-centre. 
A careful estimate of the associated uncertainty has not been carried out, but since its presence is likely to be masked by noise we introduce an uncertainty of FWHM/SNR, which is \si60\,kHz.  
A shift due to the first order Zeeman effect should not be significant here, since any residual $B$-field  will split  (or broaden) the line symmetrically, and the linearly polarised light produces equal Zeeman sub-level excitations. 

 From the data in Fig.~\ref{LineMeasurements}(a) the mean frequency for the $^{1}S_{0}\, (F=\frac{1}{2})-\,^{3}P_{1}\, (F=\frac{3}{2})$ transition   is  539\,390\,405\,433\,kHz  
 with a statistical uncertainty of 16\,kHz (compare with  the transition's natural line-width of  183.8\,kHz~\cite{Bel2012a}). 
 The overall uncertainty is dominated by the rms systematic uncertainty of 180\,kHz.   An absolute frequency for this transition can be determined from a previous measurement of the \coolingT\ transition in $^{176}$Yb and the relevant isotopic shift~\cite{Pan2009}.  The frequency is predicted to be $539\,390\,366\,\pm10$\,MHz.  We measure a frequency that is 39.4\,MHz above this value.  The frequency difference is compatible with sign errors in our frequency comb measurements, but multiple checks have been made to avoid such an oversight.  


Several tests were performed in relation to  Stark shifts on the \clockT\  transition due to the 398.9\,nm light. 
  Further assessment of the light shift has been made by probing the atoms with different levels of 398.9\,nm MOT light.  We use a cyclic scheme, where the MOT is loaded for \si0.5\,s at full intensity, then the intensity is quickly ramped down and maintained at a set intensity for a 75\,ms, during which the 578.4\,nm probe is applied. 
	 Immediately after the probe, the atom-fluorescence level is measured.   
	 Scanning the transition multiple times and curve fitting for different 398.9\,nm levels produces  the centre-frequency versus intensity  data, an example of the data is presented in Fig.~\ref{LightShift}(a).  There is a clear linear dependence with intensity, exhibiting a slope of 0.79\,MHz per unit $s_{0}$  (across seven measurements the mean slope is $0.75\pm0.11$\,MHz per unit $s_{0}$).  Also shown by the horizontal line is the same day's measurement of the centre frequency using the chopped MOT-beam approach.   A light shift of  \si160\,kHz is apparent  for this  method.  By repeating this process (across separate days), 
	  we find that the average  light shift for the MOT-beam chopping method is 170$\pm$30\,kHz.  This is applied to the clock line frequency measurements, where this approach has been used  (the 398.9\,nm intensity has varied by less than 20\,\%\ across the recording period of the measurements)~\cite{note3}. 
	   We also see the  \clockT\ line-width dependence on the MOT beam intensity in Fig.~\ref{LightShift}(b).  Interpolating to zero intensity~\cite{Cri2010} the  line-width is \si610\,kHz.  The corresponding temperature for a Doppler broadened line is 460\,\uK.   This is consistent with \Yb\ atom-cloud temperature measurements made by our group by other means\cite{Kos2014}. 
	 	   
 A Stark shift could also arise from the Zeeman-slower beam ($s_{0}\approx2.5$) if it is not sufficiently frequency detuned from the MOT-bound atoms.  For the trapped atoms in the MOT, the frequency detuning is $-7\,\Gamma$, where $\Gamma\approx182$\,rad\,s$^{-1}$ is the natural line-width of the \fastT\ transition.  Repeated measurements of the \coolingT\ transition with and without the Zeeman-slower beam showed no evidence of a light shift at 8\,kHz of resolution.    With regard to a light shift from the probe light itself, the Rabi frequency of the 578.4\,nm light associated with the \clockT\ line is \si4\,kHz.   Hence frequency shifts are not expected to be greater than this.  Line-centre frequency measurements recorded for different 578.4\,nm intensities confirmed this to be the case.  

  By differencing  the values measured in  Figs.~\ref{LineMeasurements}(a) and (b) we produce the separation frequencies for  \diffP. 
  One could avoid the comparison with the measured \clockT\ line data and rely solely on  the H-maser and prior measurements of the clock line frequency. However, by using our own clock line measurements, we limit against any bias from  the H-maser should it be grossly inaccurate. 
  The data scatter for the \diffP\ separation is similar to that of Fig.~\ref{LineMeasurements}(a), since the uncertainty of the systematic shifts  of the \coolingT\ line dominate.   We treat the uncertainty of each data point to be the rms sum of the systematic uncertainty of the \coolingT\ centre frequency, the light shift uncertainty and the statistical  uncertainty of each  \clockT\ measurement. 
   Forcing a line fit with zero gradient, the weighted mean frequency is  $21\,094\,568\,820\pm36$ (stat.) $\pm180$ (syst.)\,kHz  
with a reduced-$\chi^{2}$  of 0.36, implying that the evaluation of the systematic uncertainty is overly conservative.   However, without more effort to ensure the 555.8\,nm beams have near-flat wavefronts through the atomic beam, justifying a smaller uncertainty seems inappropriate. 

 
\section{Conclusions}

We have carried out concurrent optical frequency measurements of the $^{1}S_{0}\, (F=\frac{1}{2})-\,^{3}P_{1}\, (F=\frac{3}{2})$ and \clockT\ transitions in \Yb, and in so doing have found the frequency separation between the $^{3}P_{0}$ and  $^{3}P_{1}\, (F=\frac{3}{2})$ levels to be $21\,094\,568\,820\pm180$\,kHz.  
We note that the statistical uncertainty for the \coolingT\ measurements was below 20\,kHz; hence,  given the reproducibility, it seems plausible that the systematic uncertainty  could be reduced to a similar value. 
The knowledge of this frequency interval will be useful for those pursuing experiments with the \Yb\ \clockT\ line without access to highly accurate frequency standards, since the $^{1}S_{0}\, -\,^{3}P_{1}\, (F=\frac{3}{2})$ transition is relatively easy to make apparent. This also extends to \Ybthree, given the \clockT\ frequency measurement in \cite{Hoy2005}.   With our reported value, a wave-meter with sub-ppm accuracy should be the sole frequency measuring device required  to promptly find the \Yb\  or \Ybthree\ `clock' transitions.   We employed a hydrogen maser, but this 
should no longer  be necessary for setting up  clock-line spectroscopy in ytterbium (an alternative means of reference is described in~\cite{Hon2005a} using an $I_{2}$ cell). 
In principle, the accuracy of the \clockT\ transition frequency measurements could be improved by using a retro-reflected 578.4\,nm beam and producing the recoil doublet~\cite{Pet2008}.   The 578\,nm beam was retro-reflected with a lens-mirror cat's-eye for some of the later measurements,  but the doublet was not resolvable. 
Since the majority of the systematic uncertainty for the \diffP\ measurements lies with the \coolingT\ line, the presence of the recoil doublet is not essential. 

\section*{Acknowledgement}
This work was supported by the \ARC\ (grant LE110100054, M.E. Tobar).  J.M. is supported through an ARC Future Fellowship (FT110100392).   We are thankful to Gary Light and Steve Osborne of the UWA Physics workshop for fabricating components of the experiment.   We thank members of the ARC Centre of Excellence for Engineered Quantum Systems  for their assistance, and M.E. Tobar, S. Parker and E. Ivanov for the use of equipment.  We also thank D. Sampson and D. Lorenser from the Optical and Biomedical Engineering Laboratory, UWA, for the use of measurement instrumentation. 



\begin{thebibliography}{10}
\newcommand{\enquote}[1]{``#1''}

\bibitem{Tak2003b}
Y.~Takasu, K.~Maki, K.~Komori, T.~Takano, K.~Honda, M.~Kumakura, T.~Yabuzaki,
  and Y.~Takahashi, \enquote{Spin-singlet {Bose-Einstein} condensation of
  two-electron atoms,} Phys. Rev. Lett. \textbf{91}, 040404 (2003).

\bibitem{Fuk2007}
T.~Fukuhara, S.~Sugawa, and Y.~Takahashi, \enquote{{Bose-Einstein} condensation
  of an ytterbium isotope,} Phys. Rev. A \textbf{76}, 0516041 (2007).

\bibitem{Fuk2007a}
T.~Fukuhara, Y.~Takasu, M.~Kumakura, and Y.~Takahashi, \enquote{Degenerate
  fermi gases of ytterbium,} Phys. Rev. Lett. \textbf{98}, 030401 (2007).

\bibitem{Hin2013}
N.~Hinkley, J.~A. Sherman, N.~B. Phillips, M.~Schioppo, N.~D. Lemke, K.~Beloy,
  M.~Pizzocaro, C.~W. Oates, and A.~D. Ludlow, \enquote{An atomic clock with
  $10^{-18}$ instability,} Science \textbf{341}, 1215--1218 (2013).

\bibitem{Lem2009a}
N.~D. Lemke, A.~D. Ludlow, Z.~W. Barber, T.~M. Fortier, S.~A. Diddams,
  Y.~Jiang, S.~R. Jefferts, T.~P. Heavner, T.~E. Parker, and C.~W. Oates,
  \enquote{Spin-$1/2$ optical lattice clock,} Phys. Rev. Lett. \textbf{103},
  063001 (2009).

\bibitem{Yas2012}
M.~Yasuda, H.~Inaba, T.~Kohno, T.~Tanabe, Y.~Nakajima, K.~Hosaka, D.~Akamatsu,
  A.~Onae, T.~Suzuyama, M.~Amemiya, and F.-L. Hong, \enquote{Improved absolute
  frequency measurement of the $^{171}${Yb} optical lattice clock towards a
  candidate for the redefinition of the second,} Appl. Phys. Express
  \textbf{5}, 102401 (2012).

\bibitem{Par2013}
C.~Y. Park, D.-H. Yu, W.-K. Lee, S.~E. Park, E.~B. Kim, S.~K. Lee, J.~W. Cho,
  T.~H. Yoon, J.~Mun, S.~J. Park, T.~Y. Kwon, and S.-B. Lee, \enquote{Absolute
  frequency measurement of {$^{1}S_{0}$ ( F = 1/2)-- $^{3}P_{0}$ ( F = 1/2)}
  transition of 171 {Yb} atoms in a one-dimensional optical lattice at
  {KRISS},} Metrologia \textbf{50}, 119 (2013).

\bibitem{Tak2015a}
M.~Takamoto, I.~Ushijima, M.~Das, N.~Nemitz, T.~Ohkubo, K.~Yamanaka, N.~Ohmae,
  T.~Takano, T.~Akatsuka, A.~Yamaguchi, and H.~Katori, \enquote{Frequency
  ratios of {Sr, Yb, and Hg} based optical lattice clocks and their
  applications,} Comptes Rendus Physique \textbf{16}, 489 (2015).

\bibitem{Piz2012}
M.~Pizzocaro, G.~Costanzo, A.~Godone, F.~Levi, A.~Mura, M.~Zoppi, and
  D.~Calonico, \enquote{Realization of an ultrastable 578-nm laser for an {Yb}
  lattice clock,} IEEE Trans. Ultrason. Ferroelectr. Freq. Control \textbf{59},
  426 -- 31 (2012).

\bibitem{Nin2013}
C.~Ning, Z.~Min, C.~Hai-Qin, F.~Su, H.~Liang-Yu, Z.~Xiao-Hang, G.~Qi,
  J.~Yan-Yi, B.~Zhi-Yi, M.~Long-Sheng, and X.~Xin-Ye, \enquote{Clock-transition
  spectrum of {171Yb} atoms in a one-dimensional optical lattice,} Chin. Phys.
  B \textbf{22}, 090601 (2013).

\bibitem{Ger2010a}
F.~Gerbier and J.~Dalibard, \enquote{Gauge fields for ultracold atoms in
  optical superlattices,} New J. Phys. \textbf{12}, 033007 (2010).

\bibitem{Dal2011}
J.~Dalibard, F.~Gerbier, G.~Juzeliunas, and P.~Ohberg, \enquote{Colloquium:
  Artificial gauge potentials for neutral atoms,} Rev. Mod. Phys. \textbf{83},
  1523--43 (2011).

\bibitem{Pag2014}
G.~Pagano, M.~Mancini, G.~Cappellini, P.~Lombardi, F.~Schafer, H.~Hu, X.-J.
  Liu, J.~Catani, C.~Sias, M.~Inguscio, and L.~Fallani, \enquote{A
  one-dimensional liquid of fermions with tunable spin,} Nat. Phys.
  \textbf{10}, 198 -- 201 (2014).

\bibitem{Sca2014}
F.~Scazza, C.~Hofrichter, M.~Hoefer, P.~C. De~Groot, I.~Bloch, and S.~Foelling,
  \enquote{{Observation of two-orbital spin-exchange interactions with
  ultracold SU(N)-symmetric fermions},} Nature Physics \textbf{{10}},
  {779--784} ({2014}).

\bibitem{Non2013}
H.~Nonne, M.~Moliner, S.~Capponi, P.~Lecheminant, and K.~Totsuka,
  \enquote{Symmetry-protected topological phases of alkaline-earth cold
  fermionic atoms in one dimension,} Europhys. Lett. \textbf{102}, 37008
  (2013).

\bibitem{Hoy2005}
C.~Hoyt, Z.~Barber, C.~Oates, T.~Fortier, S.~Diddams, and L.~Hollberg,
  \enquote{Observation and absolute frequency measurements of the
  {$^{1}S_{0}-^{3}P_{0}$} optical clock transition in neutral ytterbium,} Phys.
  Rev. Lett. \textbf{95}, 083003--1 (2005).

\bibitem{Cla1979}
D.~Clark, M.~Cage, D.~Lewis, and G.~Greenlees, \enquote{Optical isotopic shifts
  and hyperfine splittings for {Yb},} Phys. Rev. A \textbf{20}, 239 (1979).

\bibitem{Wij1994}
W.~Wijngaarden and J.~Li, \enquote{Measurement of isotope shifts and hyperfine
  splittings of ytterbium by means of acousto-optic modulation,} J. Opt. Soc.
  Am. B \textbf{11}, 2163 (1994).

\bibitem{Pan2009}
K.~Pandey, A.~K. Singh, P.~V.~K. Kumar, M.~V. Suryanarayana, and V.~Natarajan,
  \enquote{Isotope shifts and hyperfine structure in the 555.8-nm
  ${^{1}S}_{0}\ensuremath{\rightarrow}{^{3}P}_{1}$ line of {Yb},} Phys. Rev. A
  \textbf{80}, 022518 (2009).

\bibitem{Dal2008}
A.~Daley, M.~Boyd, J.~Ye, and P.~Zoller, \enquote{Quantum computing with
  alkaline-earth-metal atoms,} Phys. Rev. Lett. \textbf{101}, 170504 (2008).

\bibitem{Shi2009}
K.~Shibata, S.~Kato, A.~Yamaguchi, S.~Uetake, and Y.~Takahashi, \enquote{A
  scalable quantum computer with ultranarrow optical transition of ultracold
  neutral atoms in an optical lattice,} Appl. Phys. B \textbf{97}, 753--8
  (2009).

\bibitem{Dal2011a}
A.~J. Daley, \enquote{Quantum computing and quantum simulation with {group-II}
  atoms,} Quantum Information Processing \textbf{10}, 865--884 (2011).

\bibitem{Wol2007}
P.~Wolf, P.~Lemonde, A.~Lambrecht, S.~Bize, A.~Landragin, and A.~Clairon,
  \enquote{From optical lattice clocks to the measurement of forces in the
  {Casimir} regime,} Phys. Rev. A \textbf{75}, 063608 (2007).

\bibitem{Kos2014}
N.~Kostylev, E.~Ivanov, M.~Tobar, and J.~J. McFerran, \enquote{{Sub-Doppler
  cooling of ytterbium with the $^{1}S_{0}-^{1}P_{1}$ transition including
  $^{171}$Yb (I=1/2)},} J. Opt. Soc. Am. B \textbf{31}, 1614 (2014).

\bibitem{Kos2015}
N.~Kostylev, C.~Locke, M.~Tobar, and J.~McFerran, \enquote{Spectroscopy and
  laser cooling on the $^{1}s_{0}--\,^{3}p_{1}$ line in yb via an
  injection-locked diode laser at 1111.6\,nm,} Applied Physics B \textbf{118},
  517--525 (2015).

\bibitem{Han1980}
T.~H\"{a}nsch and B.~Couillaud, \enquote{Laser frequency stabilization by
  polarization spectroscopy of a reflecting reference cavity,} Optics
  Communications \textbf{35}, 441 -- 444 (1980).

\bibitem{note1}
Partly limited by a 20\,m separation between the cavity and the laser.

\bibitem{Nic2003}
J.~Nicholson, M.~Fan, P.~Wisk, J.~Fleming, F.~DiMarcello, E.~Monberg, A.~Yabon,
  C.~Jorgensen, and T.~Veng, \enquote{All-fiber, octave-spanning
  supercontinuum,} Opt. Lett. \textbf{28}, 643 (2003).

\bibitem{Num2004}
K.~Numata, A.~Kemery, and J.~Camp, \enquote{Thermal-noise limit in the
  frequency stabilization of lasers with rigid cavities,} Phys. Rev. Lett.
  \textbf{93}, 250602--1 (2004).

\bibitem{Not2006}
M.~Notcutt, L.-S. Ma, A.~D. Ludlow, S.~M. Foreman, J.~Ye, and J.~L. Hall,
  \enquote{Contribution of thermal noise to frequency stability of rigid
  optical cavity via hertz-linewidth lasers,} Physical Review A \textbf{73},
  031804 (2006).

\bibitem{McF2012b}
J.~J. McFerran, D.~V. Magalhaes, C.~Mandache, J.~Millo, W.~Zhang, Y.~L. Coq,
  G.~Santarelli, and S.~Bize, \enquote{{Laser locking to the $^{199}$Hg
  \clockTdash\ clock transition with $5.4\times 10^{-15}/\sqrt{\tau}$
  fractional frequency instability},} Opt. Lett. \textbf{37}, 3477--3479
  (2012).

\bibitem{Hon2005a}
T.~Hong, C.~Cramer, E.~Cook, W.~Nagourney, and E.~Fortson, \enquote{Observation
  of the {$^{1}S_{0}-^{3}P_{0}$} transition in atomic ytterbium for optical
  clocks and qubit arrays,} Opt. Lett. \textbf{30}, 2644 (2005).

\bibitem{Dem2008}
W.~Demtr\"{o}der, \emph{Laser Spectroscopy, Experimental Techniques}, vol.~2
  (Springer, 2008), 4th ed.

\bibitem{Bel2012a}
K.~Beloy, J.~Sherman, N.~Lemke, N.~Hinkley, C.~Oates, and A.~Ludlow,
  \enquote{Determination of the {5d6s$^{3}D_{1}$} state lifetime and
  blackbody-radiation clock shift in {Yb},} Phys. Rev. A \textbf{86}, 051404
  (2012).

\bibitem{note2}
The hydrogen-maser was last calibrated and corrected to agree with UTC-Sydney
  in November 2011 to $8\times10^{-13}$.

\bibitem{Bor1976}
C.~J. Bord\'e, J.~L. Hall, C.~V. Kunasz, and D.~G. Hummer, \enquote{Saturated
  absorption line shape: Calculation of the transit-time broadening by a
  perturbation approach,} Phys. Rev. A \textbf{14}, 236 (1976).

\bibitem{Hal1976a}
J.~Hall and C.~Borde, \enquote{Shift and broadening of saturated absorption
  resonances due to curvature of the laser wave fronts,} Appl. Phys. Lett.
  \textbf{29}, 788 (1976).

\bibitem{note3}
Interleaving the 578.4\,nm and 398.9\,nm light via AOMs is a more appropriate
  method for carrying out the clock line spectroscopy, but technical issues
  prevented its use for much of the data recording period.

\bibitem{Cri2010}
M.~Cristiani, T.~Valenzuela, H.~Gothe, and J.~Eschner, \enquote{Fast
  nondestructive temperature measurement of two-electron atoms in a
  magneto-optical trap,} Phys. Rev. A \textbf{81}, 063416 (2010).

\bibitem{Pet2008}
M.~Petersen, R.~Chicireanu, S.~T. Dawkins, D.~V. Magalhaes, C.~Mandache, Y.~L.
  Coq, A.~Clairon, and S.~Bize, \enquote{Doppler-free spectroscopy of the
  $^{1}${S}$_{0}$-$^{3}${P}$_{0}$ optical clock transition in laser-cooled
  fermionic isotopes of neutral mercury,} Phys. Rev. Lett. \textbf{101}, 183004
  (2008).

\end{thebibliography}
%


\end{document}